\documentclass[journal,twocolumn]{IEEEtran}

\IEEEoverridecommandlockouts                              
\usepackage{cite}
\usepackage{subcaption}
\usepackage{graphicx}
\usepackage{amsmath}
\usepackage{amssymb}
\usepackage[normalem]{ulem}
\usepackage[ruled,vlined]{algorithm2e}
\usepackage{framed}
\usepackage[table]{xcolor}
\usepackage{array}
\usepackage{booktabs, tabularx}
\usepackage{multirow}
\usepackage{bigints}
\usepackage{times}

\usepackage[]{algorithm2e}
\usepackage{bm}  
\usepackage{color}
\usepackage{ulem}
\graphicspath{ {images/} }

\usepackage{balance}

\newcommand{\bmsf}[1]{\bm{\mathsf{#1}}} 

\newcommand{\yudel}[1]{}
 
\usepackage{float}
\usepackage{caption}
\usepackage{subcaption}

\title{Bridging the Security Gap:\\ Lessons from 5G and What 6G Should Do Better}
\author{
\IEEEauthorblockN{Isabella D. Lutz and Matthew C. Valenti} \\
West Virginia University, Morgantown, WV, USA. \\
}

\begin{document}

\maketitle
\thispagestyle{empty}
\pagestyle{empty}

\begin{abstract}
The security requirements for future 6G mobile networks are anticipated to be significantly more complex and demanding than those of 5G. This increase stems from several factors: the proliferation of massive machine-type communications will dramatically increase the density of devices competing for network access; secure ultra-reliable low-latency communication will impose stringent requirements on security, latency, and reliability; and the widespread deployment of small cells and non-terrestrial networks, including satellite mega-constellations, will result in more frequent handovers. This paper provides a set of security recommendations for 6G networks, with a particular focus on access and handover procedures, which often lack encryption and integrity protection, making them more vulnerable to exploitation. Since 6G is expected to be a backward-compatible extension of 5G, and given that secure systems cannot be effectively designed without a clear understanding of their goals, it is imperative to first evaluate the limitations of the current generation. To this end, the paper begins by reviewing existing 5G access and authentication mechanisms, highlighting several critical vulnerabilities in these procedures. It then examines potential 6G challenges and concludes with actionable recommendations to enhance the security, resilience, and robustness of 6G access and handover mechanisms.




\end{abstract}

\vspace{-0.2cm}

\section{Introduction}
\label{intro}

The sixth generation (6G) of mobile networks is anticipated to revolutionize wireless communication by offering unprecedented levels of connectivity, speed, and intelligence while placing more emphasis on efficiency as quantified by cost (bits per dollar) and energy (bits per Joule) \cite{andrews2024}. Building upon the foundational advances of 5G, 6G networks will enable new applications across sectors such as autonomous transportation, remote surgery, extended reality, and massive Internet of Things (IoT) ecosystems. However, these advancements come with significant challenges, particularly in the realms of access and authentication procedures, which must evolve to support the heightened demands of a 6G environment. As these networks aim to provide secure ultra-reliable low-latency communication (sURLLC) and seamless connectivity, ensuring secure and efficient access control becomes increasingly critical.   

One of the most pressing challenges in 6G networks will be managing the surge in device density driven by massive machine-type communications (mMTC). With millions (or even billions) of devices, including sensors, vehicles, and wearables, competing for network resources, traditional access and authentication mechanisms may become inefficient or inadequate. The need for secure and reliable communication in environments with high device densities will push the boundaries of what existing technologies can support, necessitating innovative solutions that can deliver high performance without compromising security.

Furthermore, the rise of non-terrestrial networks (NTN) and the proliferation of small cell deployments will lead to a communication landscape characterized by frequent handovers and complex mobility patterns \cite{guidotti2024}. Satellite mega-constellations and aerial networks, for instance, introduce new dimensions of mobility, requiring handoff procedures that are fast, reliable, and secure. As users and devices seamlessly transition between terrestrial and non-terrestrial infrastructures, access control mechanisms must be highly adaptable and capable of maintaining performance without interruptions or vulnerabilities. This evolution demands a reassessment of conventional authentication techniques, which may struggle under these conditions.

Another transformative aspect of 6G will be the widespread integration of artificial intelligence (AI) and machine learning (ML) to enhance network operations \cite{porambage2021roadmap}. AI capabilities will not only optimize resource allocation and predict traffic patterns but also enable adaptive and context-aware access and authentication mechanisms. By leveraging AI, 6G networks can dynamically adjust security measures based on real-time risk assessments, user behavior, and network conditions. Despite these potential benefits, integrating AI into security frameworks raises concerns around privacy, model robustness, and susceptibility to adversarial attacks, all of which must be addressed to ensure reliable performance \cite{kazmi2023security}.


In this paper, we provide a set of recommendations, which could be used to form a roadmap, for developing secure procedures for access and authentication in highly mobile 6G networks.  We begin by reviewing the current state of access and authentication in 5G, with a focus on their limitations as highlighted by a set of known exploits that we review. Having identified these vulnerabilities, we provide some insights into how 6G could be better architected to avoid these issues.   A key contribution of this paper is that its recommendations are built upon a strong foundation of understanding 5G security procedures and their vulnerabilities.  This is in contrast with other papers on 6G security, such as \cite{porambage2021roadmap, kazmi2023security,hakeem2023security}, which are far less specific about actual underlying security procedures.   

The paper is organized as follows. Sec.~\ref{section2} provides an overview of security  procedures used in the 5G NR system. Sec.~\ref{section3} identifies several known security vulnerabilities of 5G systems that may occur during access, authentication, and handover. Sec.~\ref{section4} provides a set of recommendations for 6G, offering several solutions to the issues identified in the paper. Finally, Sec.~\ref{conclusion} concludes the paper.

\section{Overview of 5G Security Procedures}
\label{section2}

This section provides an overview of access and authentication procedures used in the 5G New Radio (NR) system.  
In particular, we walk through these procedures starting from when a mobile first powers up leading to when it is first authenticated.  We then proceed to discuss handover procedures.

\subsection{Network Architecture}
\begin{figure}[t]
\centering
\includegraphics[width=0.43\textwidth]{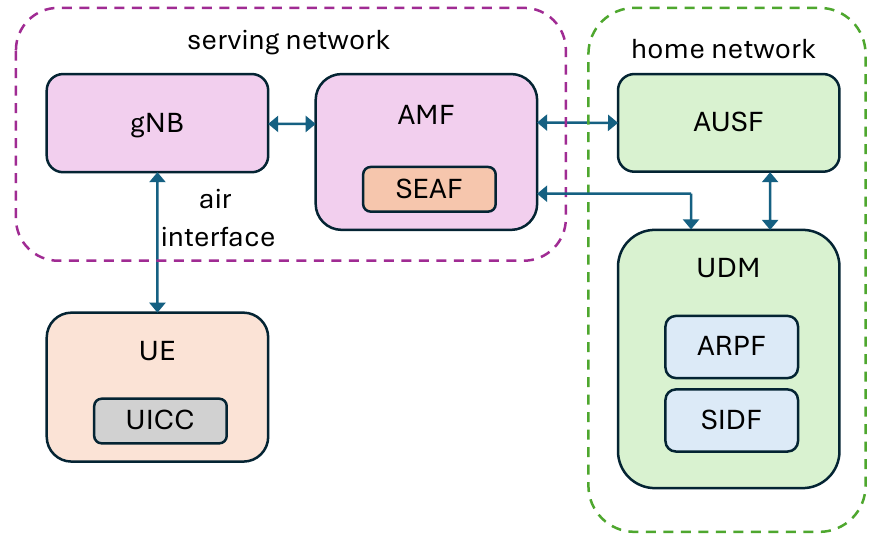}
\caption{{\small Simplified 5G network architecture.}}
\vspace{-0.25cm}
\label{fig:net}
\end{figure}

Before walking through the steps required for a mobile to access the network and activate its security context, we first describe the architecture of a typical 5G NR network, focusing on only those network components that are required for these procedures.  Fig. \ref{fig:net} shows a simplified 5G NR network with only the components needed for access and authentication.    

The \emph{Radio Access Network} is referred to as NG-RAN and comprises \emph{base stations} known as gNB that connected over an \emph{air interface} to mobiles known as \emph{user equipment} (UE) \cite{cox2021book,dahlman2021book}.  The UE contains a \emph{Universal Integrated Circuit Card} (UICC) \footnote{Most modern phones, such as the iPhone 14 and above, have an \emph{Embedded SIM} (eSIM) instead of a physical UICC.  As the eSIM is fully compliant with UICC standards, we use the term \emph{UICC} to encompass both possibilities.}, which acts as a platform to run multiple applications including the \emph{Universal Subscriber Identity Module} (USIM).  The USIM includes important information related to the subscriber's network, authentication keys, and its identity. 

The UE is connected via the NG-RAN to an \emph{Access and Mobility Management Function} (AMF), which is the component of the serving (or visited) 5G core network that handles access management, including management of cryptographic keys, and mobility management \cite{3GPP.TS.23.501}. Co-located with the AMF is the \emph{Security Anchor Function} (SEAF), which manages the authentication procedures of the \emph{Serving Network} (SN). The AMF is connected to, and works closely with, the \emph{Unified Data Management} (UDM), which resides in the UE's \emph{Home Network} (HN) and includes a subscriber database and an \emph{Authentication Credential Repository and Processing Function} (ARPF), which stores the most important security keys and carries out security-related calculations.  The UDM also contains a \emph{Subscription Identifier De-concealing Function} (SIDF), which is responsible for deconcealing the UE's identity \cite{3GPP.TS.33.501}.  Additionally, the UDM is connected to an \emph{Authentication Server Function} (AUSF), which resides in the home network and manages authentication procedures. 


\subsection{Cell and Network Acquisition}
\label{acquisition}
Upon power-up, a 5G NR UE begins by initializing its hardware and software components, including its radio and protocol stack. The UE retrieves subscription information from its USIM, including the \emph{Home Public Land Mobile Network} (HPLMN), a list of preferred PLMNs used when roaming outside its HPLMN, and a list of forbidden PLMNs \cite{cox2021book}. This information guides the UE in prioritizing which networks to connect to during its search. The UE then scans frequencies based on a predefined frequency raster that ensures efficient and systematic scanning by defining evenly spaced frequencies within supported bands. 
The search may be accelerated by using a list of previously connected cells and frequencies, but if no such list is available, a full band scan will be performed over the bands that the UE is configured to support.  

As the UE scans, it searches for \emph{Synchronization Signal Blocks} (SSBs) broadcast by nearby gNBs \cite{dahlman2021book}. Each SSB consists of a \emph{Primary Synchronization Signal} (PSS) and a \emph{Secondary Synchronization Signal} (SSS), used for cell identification and timing synchronization, along with the \emph{Physical Broadcast Channel} (PBCH) \cite{3GPP.TS.38.211}. The PBCH carries the \emph{Master Information Block} (MIB), which is a \emph{polar coded} message that provides essential cell configuration parameters, including the \emph{system frame number} (SFN) and the time-frequency location of \emph{CORESET\#0} (Control Resource Set 0) \cite{3GPP.TS.38.331}. CORESET\#0 is a time-frequency region containing the \emph{Physical Downlink Control Channel} (PDCCH) that carries the scheduling information for \emph{System Information Block 1} (SIB1).  SIB1 is a message sent over the \emph{Physical Downlink Shared Channel} (PDSCH) containing the \emph{Remaining Minimum System Information} (RMSI) that the UE needs to connect to the network.   The number of SSBs transmitted per time interval depends on the network's beamforming configuration; in cases of highly directional beamforming, multiple SSBs are transmitted over different directions to ensure coverage, with up to 64 beams supported in mmWave bands. 

Once SIB1 is decoded, the UE gains access to detailed system information such as the PLMN identity and cell access restrictions, enabling it to proceed with registration and initial connection setup \cite{3GPP.TS.38.331}.  If the PLMN matches the HPLMN or preferred PLMN stored on the USIM, then the UE will enter RRC\_IDLE mode and begin \emph{camping} on the cell.  While camped it will listen for incoming notifications over the \emph{Paging Channel} (PCH) according to its \emph{Discontinuous Reception} (DRX) paging cycle.  If the UE has moved into a new \emph{tracking area} (TA) -- which it can determine by comparing the \emph{tracking area code} (TAC) carried by SIB1 to its current \emph{tracking area list} (TAL) or upon expiry of a timer  -- then the UE will initiate a \emph{mobility registration update} \cite{3GPP.TS.23.501}. 

\subsection{Registration}

The UE will perform an \emph{random access} (RA) procedure for any of the following reasons (called \emph{establishment causes}): the UE has data to transmit; it is notified by the PCH that it will receive incoming data; or the UE needs to perform a signaling procedure such as a registration update or a request for \emph{Other System Information} (OSI), which are on-demand System Information Blocks that follow SIB1 \cite{3GPP.TS.38.331}. The RA procedure involves the UE selecting a \emph{random access preamble} from a pool of preambles configured by SIB1 \cite{3GPP.TS.38.321}.  The randomly selected preamble is sent by the UE over the \emph{Physical Random Access Channel} (PRACH) during the UE's preconfigured \emph{random access occasions} \cite{3GPP.TS.38.211}.  The base station will respond with a \emph{Random Access Response} (RAR) containing the index of the received preamble, a \emph{Timing Advance} (TA) used to correct uplink timing, an uplink grant, and a temporary identity known as \emph{TC-RNTI} \cite{3GPP.TS.38.321}.  

Using the given uplink grant and TA, the UE will respond with \emph{msg3} containing an \emph{RRC Setup Request} message, which includes the UE's unique identifier and the establishment cause \cite{3GPP.TS.38.331}.  Finally, the gNB responds with a \emph{contention resolution} (CR) message, which will be either a \emph{RRC Setup} message or a \emph{RRC Reject} message \cite{3GPP.TS.38.331}.  The unique identifier of a UE will be present in the CR message, and contention is resolved if it matches the identifier sent in msg3.  Otherwise, a collision may have occurred due to two UEs selecting the same preamble for transmission during the same random access occasion.  Upon a successful contention resolution and reception of RRC Setup, the UE will be assigned a C-RNTI (usually by elevating the TC-RNTI) that will be used to address subsequent \emph{Downlink Control Information} (DCI) sent by the gNB over the PDCCH.   The UE is now in the RRC\_CONNECTED state.   The UE then acknowledges its configuration by sending a \emph{RRC Setup Complete} message, which includes a \emph{Registration Request} message and the globally unique identity of the AMF with which the UE was previously registered.   

The gNB proceeds to select an AMF, using the one that the UE was previously connected to if possible.  The gNB sends an \emph{Initial UE Message}, which encapsulates the UE's Registration Request, to the selected AMF to begin the registration procedure \cite{3GPP.TS.23.501}.  The message also includes a \emph{Non-Access Stratum} (NAS) \emph{Key Set Identifier} (KSI), which identifies the security context including the set of security keys most recently used by the UE \cite{3GPP.TS.33.501}.  If the UE was most recently registered with the AMF, then the AMF will retrieve a stored copy of the UE's context, otherwise it will attempt to contact the previous AMF through a \emph{UEContextTransfer} operation to obtain the context from the old AMF.   If the UE does not have a valid security context, e.g., it is performing initial registration with a new and unknown AMF, then the NAS KSI is set to a reserved value of all ones and a full authentication will follow.    On the other hand, if the UE does have a valid security context, then it will use the NAS KSI to determine the key set and utilize the associated cryptographic keys to cipher and integrity-protect subsequent NAS messages.  Importantly, when the UE lacks a valid security context, the Registration Request and other initial messages are sent in the clear, without encryption. Moreover, even if the UE has a valid security context, certain fields including fields related to AMF selection and the NAS KSI itself, are sent in the clear.

\subsection{Authentication and Key Generation}
\label{aka}
Security at the access level relies on a user-specific key $K$, which is securely stored both at the UE in its UICC and in the HN at the ARPF \cite{cox2021book,3GPP.TS.33.501}.  From the shared key $K$, a hierarchy of other keys are derived.   When the AMF receives a Registration Request, it selects an AUSF in the mobile's home network and asks it to authenticate the mobile.  When it does this, it passes the subscriber's identity, which is concealed as a \emph{Subscription Concealed Identifier} (SUCI), which is a public key encrypted version of the \emph{Subscription Permanent Identifier} (SUPI).  Concealing the UE's identifier as a SUCI is an important measure to prevent the mobile from being cloned by an adversary. The SUCI is decoded by the SIDF co-located at the UDM by using a public key stored in the home network's ARPF. After retrieving the SUPI, the home network chooses an authentication method, which may be either \emph{5G AKA} or \emph{EAP-AKA'}.  


The network and UE then initiate an \emph{Authentication and Key Agreement} (AKA) protocol  \cite{cox2021book,3GPP.TS.33.501}.  The AKA protocol relies on a counter called the \emph{sequence number} (SQN) to track successful handshakes and provide replay protection. Ideally, the SQN at the UE and network are the same, but they may become unsynchronized due to message loss. To handle the possibility of mismatched sequence number, we use the variable $\mathsf{SQN_{UE}}$ to denote the sequence number stored at the UE and the variable $\mathsf{SQN_{HN}}$ to denote the sequence number stored at the home network's ARPF. 

After identifying the UE, the ARPF retrieves the user-specific key ($K$) and generates an \emph{authentication challenge vector} consisting of four elements \cite{3GPP.TS.33.501}:
\begin{enumerate}
    \item {$\bmsf{RAND}$}: A random nonce used as an authentication challenge to the mobile.
    \item {$\bmsf{AUTN}$}: An authentication token that demonstrates the network possesses the correct $K$.  It includes an \emph{Authentication Management Field} ($\mathsf{AMF}$), a concealed version of $\mathsf{SQN_{HN}}$, and a \emph{Message Authentication Code} (MAC) formed by concatenating $\mathsf{SQN_{HN}}$, $\mathsf{RAND}$, and $\mathsf{AMF}$, then encrypting using $K$.
    \item {$\bmsf{XRES*}$}: The expected response to the challenge, which can only be computed by a device with the correct $K$.
    \item {$\bmsf{K_{AUSF}}$}: The anchor key used by the home network.  It is derived from $K$,  $\mathsf{RAND}$, and the serving network name.
\end{enumerate}
The ARPF sends this authentication vector back to the AUSF, along with the SUPI if retrieved. 
The AUSF stores $\mathsf{XRES*}$ 
and computes $\mathsf{HXRES*}$, which is a hashed version used by the serving network. The AUSF also generates the serving network’s anchor key \( K_\mathsf{SEAF} \) which if forwards to the AMF along with $\mathsf{HXRES*}$, $\mathsf{RAND}$, and  $\mathsf{AUTN}$ 
The AMF then sends  $\mathsf{RAND}$ and  $\mathsf{AUTN}$ to the mobile.

On the mobile side, $\mathsf{RAND}$ and  $\mathsf{AUTN}$ are used by the USIM application to verify that the network possesses the correct \( K \) and that the sequence number is valid. If successful, the USIM computes the response \( \mathsf{RES} \) and passes it to the mobile equipment along with two cryptographic keys: a \emph{cipher key} \( \mathsf{CK} \) and an \emph{integrity key} \( \mathsf{IK} \). The mobile equipment uses \( \mathsf{RES} \), the serving network name, and these keys to calculate the 5G-specific authentication response \( \mathsf{RES*} \) and derives other keys in the hierarchy denoted \( K_\mathsf{AUSF} \), \( K_\mathsf{SEAF} \), and \( K_\mathsf{AMF} \).

The mobile sends \( \mathsf{RES*} \) to the AMF, which hashes it to compute \( \mathsf{HRES*} \) and compares it with \( \mathsf{HXRES*} \). If they match, the mobile is authenticated. The AMF confirms the authentication with the AUSF, including \( \mathsf{RES*} \). The AUSF validates \( \mathsf{RES*} \) against \( \mathsf{XRES*} \), ensuring the mobile and serving network are genuine. The AUSF acknowledges the AMF and includes any permanent identity retrieved earlier. The AMF then computes \( K_\mathsf{AMF} \), which is the key in the hierarchy associated with the AMF.    The AUSF then notifies the UDM of the successful authentication.  




At this point the NAS keys have been activated, but a set of Access Stratum (AS) keys are still needed for communication between the gNB and UE.   The AS keys are hierarchical and all are based on a key called $K_\mathsf{gNB}$, which is derived from $K_\mathsf{AMF}$, the identity of the gNB, and the bearer frequency of the serving cell.  As these quantities are know by both the gNB and UE, both can derive the same key.

\subsection{Security During Handover and Mobility}
\label{handover}
The process of moving a UE from one cell to another cell is called \emph{handover}.  Handover is facilitated by the UE performing periodic measurements of surrounding base stations and sending a \emph{Measurement Report} (MR) back to its serving gNB when certain conditions are triggered  \cite{cox2021book,3GPP.TS.38.331}.  The measurement configuration is provided to the UE at the time it is put into RRC\_CONNECTED.  The configuration tells the UE which frequencies and cells to monitor, thresholds for triggering reports, and other reporting details.  The UE typically monitors each nearby gNB by listening for its SSB, including its PSS and SSS, using those signals to compute signal measurement values \cite{cox2021book}. 
While there are several kinds of triggers, a 
typical trigger would be when a monitored neighboring cell becomes sufficiently better than the serving cell by a specified margin.  When the conditions for a trigger are met, the UE sends a \emph{RRC Measurement Report}. 

Once the gNB receives the measurement report, it determines if a handover is necessary.  If indeed a handover is required, the serving gNB selects a new (target) cell and initiates a handover procedure by first sending a \emph{Handover Request} message to the target gNB, which includes the UE's context (including its security context).  If the target gNB accepts the handover request, it returns a \emph{Handover Request Acknowledge} message to the old (source) gNB, and then the source gNB sends a \emph{RRC Reconfiguration} message to the UE with information about the handover.  The UE then attempts a random access procedure with the target gNB, possibly using a reserved random access preamble that will avoid contention.  Once the UE successfully connects to the target gNB, the UE sends a \emph{RRC Reconfiguration Complete} message to the target gNB.

During the procedure, the source gNB generates a new AS security key $K_\mathsf{gNB}^*$ which is sent to the target gNB in the \emph{Handover Request} message  \cite{3GPP.TS.33.501}.  If the same AMF will be used after handover, the new key is derived by the source gNB by using the current AS security key $K_\mathsf{gNB}$, the identity of the target gNB, and a \emph{Next Hop Chaining Counter} (NHCC) that prevents key reuse across multiple handovers.  The UE is able to simultaneously derive the same $K_\mathsf{gNB}^*$ using the same inputs.  Upon successful handover, the NHCC is incremented.  
While the AS keys have changed, the system maintains the same NAS keys unless a new AMF is used.



\section{Security Vulnerabilities in 5G}
\label{section3}
This section identifies several known security vulnerabilities in 5G systems that may occur during access, authentication, and handover, which are states when the UE is most vulnerable. These include attacks on the cell acquisition procedure by spoofing a full or partial SSB, vulnerabilities in the synchronization process of the 5G AKA protocol, 
and attacks that exploit the handover process. 

\subsection{Cell Acquisition Attacks} 
\label{acquisition_attack}
As described in Section \ref{acquisition}, the cell acquisition process involves having the UE search for the SSBs transmitted by nearby base stations. The predictable and unprotected nature of SSBs make them vulnerable to manipulation.  This vulnerability could be exploited by an attacker that uses a radio device, such as a \emph{software defined radio} (SDR), to transmit a spoofed SSB containing a bogus MIB. Such an attack was proposed in \cite{ludant2021sigunder} and is called the $\mathsf{SigUnder}$ attack.  

In $\mathsf{SigUnder}$, the attacking SDR acquires timing information and the identity of the victim gNB by reading the PSS and SSS within its SSB.   It then creates a spoofed version of that SSB, which it transmits towards the victim UE.   The spoofed version may differ from the genuine version by changing values of some of the fields in the MIB.  For instance, by changing the value of the $\mathsf{CellBarred}$ flag, the spoofed SSB would now indicate that the UE is barred from connecting to the gNB.   A straightforward implementation of this attack would transmit the entire spoofed SSB, but it would need to be transmitted at high power in order for the UE to capture it in favor of the genuine SSB.  

In a clever twist, \cite{ludant2021sigunder} shows how it is possible to change just a subset of the bits in the SSB, namely those polar code bits that would have changed due to the changed MIB bits, and then only transmit the changed portion of the SSB, potentially along with the accompanying \emph{Demodulation Reference Symbols} (DM-RS).  This has the benefit of requiring less transmit power, since only the changed bits need to be transmitted, thereby making it more difficult to detect the attack, but requires the attacker to carefully adjust its timing, carrier frequency, and phase offset.  
A mitigation is proposed in \cite{ludant2021sigunder} called $\mathsf{SICUnder}$, but it requires the use of \emph{successive interference cancellation}, which involves extensive signal processing. 
When carried out successfully, $\mathsf{SigUnder}$ can cause the UE to fail synchronization with the base station (gNB), resulting in disruption to critical 5G functions (e.g. cell selection, reselection, and handovers).



\subsection{Key Synchronization Attacks}
The AKA protocol is used to enforce the protection of identity, authentication, and verification between users and the network. As described in Section \ref{aka}, the SQN tracks the successful handshakes of a UE connecting to a serving network, forwarding an authentication request to the home network. The AKA protocol facilitates this need for mutual authentication to allow a subscriber access to their registered network securely. However, despite its improvements from previous network generations, vulnerabilities such as IP address spoofing, replay, \emph{Man-in-the-Middle} (MiM), denial of service, and IMSI-catchers persist. For instance, a formal analysis in \cite{basin2018formal},
uses the \emph{Tamarin} verification tool
to show that the 5G-AKA protocol is vulnerable to MiM attacks, replay attacks, synchronization failures, and location tracking.  

Moreover, as shown in 
\cite{cheng2022tracking},  an attacker can steal personal information from a subscribed user by deliberately triggering a synchronization failure. Within the 5G-AKA protocol, the SQN assigned to the UE and HN is synchronized, starting at the same value and incrementing 
after each successful handshake. For example, the HN increments $\mathsf{SQN_{HN}}$ after confirming the subscriber's identity. Then, the UE increments $\mathsf{SQN_{UE}}$ after verifying authentication. The vulnerability arises from failed synchronization when an outdated authentication request is reused, causing a mismatch between $\mathsf{SQN_{HN}}$ and  $\mathsf{SQN_{UE}}$. In the failure scenario,  at initialization of time=1, $\mathsf{SQN_{HN}}=\mathsf{SQN_{UE}}=1$, and finishes successfully with values $\mathsf{SQN_{HN}}=\mathsf{SQN_{UE}}=2$. However, when the AKA protocol is executed again at $t=i$, if an outdated authentication request is received instead of the expected values for $t=i$, a mismatch occurs because the reused request corresponds to a lower $\mathsf{SQN_{HN}}$ leading to failed authentication. Preventing outdated or replayed authentication requests in 5G-AKA is crucial, as attackers can exploit $\mathsf{SQN_{HN}}$ discrepancies to track subscriber patterns, such as visit frequency and duration at locations, ultimately exposing sensitive personal information.

\subsection{Handover Attacks}
\label{handover_attack}
As discussed in Section \ref{handover}, the handover procedure relies on the UE taking measurements of surrounding base stations and reporting these measurements to the source base station in a Measurement Report. These measurements are based on the PSS and SSS, which are synchronization signals that are unencrypted deterministic functions of the \emph{Physical Cell ID} (PCI) of the gNB, and therefore vulnerable for exploitation.

For example, as described in \cite{bitsikas2021handover}, an attacker can exploit this procedure to spoof a base station and disrupt the handover process. 
To execute this exploit, the attacker begins by conducting reconnaissance to gather system information about gNBs in the targeted area. This is accomplished by detecting each gNB's SSB. The PSS and SSS within the SSB reveal the gNB’s PCI, while the BCH provides the MIB. Additionally, the attacker can decode SIB1 for each gNB to obtain further critical network information, such as the \emph{Tracking Area Identifier} (TAI).

Based on this information, the attacker builds a replicated model of the network structure, revealing vulnerable base stations, and selects a victim source gNB to exploit and a victim target gNB to spoof. Ideally, the spoofed target gNB should be one listed in the \emph{Neighbor Cell List} (NCL) provided by the victim source gNB to the UEs it serves.
The attacker can then create a spoofed gNB using an SDR and open-source software. The spoofed gNB mimics the victim target gNB by replaying its PSS and SSS signals obtained during the reconnaissance phase.   The key to the exploit is transmitting the PSS and SSS at a power level high enough to trigger a measurement report from the victim UE.  The UE includes the identity of the spoofed gNB in its Measurement Report and sends it to the source gNB over their encrypted connection. Despite the report being based on false measurements, the source gNB accepts it as valid because it is transmitted over the secure link. If the reported signal strength of the spoofed gNB is sufficiently high, the source gNB initiates the handover procedure, as described in Section \ref{handover}, causing the UE to detach from the source gNB.

As part of the handover procedure, the UE attempts a RA procedure with the spoofed gNB. If the spoofed gNB is able to conduct a MiM attack and receive network-side messages, such as the Handover Request sent from the source gNB to the legitimate target gNB, it can then establish active communication with the UE. Otherwise, the spoofed gNB can still execute a \emph{Denial of Service} (DoS) attack by continuously causing the RA procedure to fail. As the UE perceives the spoofed gNB as a legitimate cell with the strongest signal, it may remain camped on the spoofed gNB for an extended period. During this time, the UE is unable to receive Paging Messages, effectively cutting off its communication with the network.

\section{6G Security Issues and Opportunities}
\label{section4}

This section provides specific recommendations for handling security in 6G systems.  The recommendations are informed by the current literature and provide solutions to some of the vulnerabilities that were identified in Section III.  

\subsection{6G Cell Acquisition}

The specific cell acquisition attacks outlined in Section \ref{acquisition_attack} and handover attacks outlined in Section \ref{handover_attack} are only possible because the predictability of the SSB
make it easy to stand up a fake base station that replays these signals, possibly with malicious modifications.  Moreover, there is no authentication procedure to identify such cloned signals.   

A possible solution to prevent the handover attack of Section \ref{handover_attack} is to embed a supplemental authentication signal into the synchronization signals.  The authentication signal could be a pseudorandom sequence known only to the legitimate network.  The UE would then be required to include the sequence (or an indicator of it) with its measurement report, and the network would only execute the handover if the correct sequence was sent.  To prevent a replay attack, the pseudorandom sequence would have an extremely long periodicity.   The sequence could either be embedded within the existing PSS and SSS structure, or a new \emph{ternary synchronization signal} (TSS) could be used. 

Preventing the acquisition attacks of Section \ref{acquisition_attack} is more challenging as the UE may only be interacting with the rogue base station.   The additional authentication signal described above could still be used, but there would be a need to make the UE aware of the signal a priori, rather than merely having it send the signal back to the network.  Another option may be to  provide strong integrity protection of the Minimum System Information (MSI) conveyed by the MIB and SIB1.  Implementation details on how this would work is left as an open problem.  


\subsection{6G-AKA Protocol}
In contrast to the vulnerabilities identified in the 5G-AKA protocol, it is anticipated that the 6G-AKA protocol will address these issues by creating a more sophisticated network architecture. The 6G network is projected to be highly cloud-based and incorporate open, programmable networking technologies. This creates the need for robust and adaptable authentication mechanisms to secure its dynamic environment \cite{hakeem2023security}. To achieve this, the 6G-AKA protocol introduces innovations such as quantum-safe cryptography and physical layer security to protect against advanced threats, including impersonation attacks.

A critical improvement in 6G-AKA is its ability to manage authentication in a deep-sliced, programmable network infrastructure, which is essential for supporting cross-slice communications. For determining authentication in these types of scenarios, the protocol must ensure clarity on which component -- SEAF or AUSF —- is responsible \cite{hakeem2023security}. Furthermore, 6G-AKA is expected to incorporate a new user identity management approach that builds on 5G's concept of a unified authentication platform for open-access networks. These advancements defend against evolving cybersecurity threats and enhance overall network access management.

\subsection{6G Handover}
The 6G network is envisioned to have a multi-layered and dynamic architecture to support seamless connectivity and high mobility. This includes an integrated environment with terrestrial and satellite domains, along with multiple mobile entities \cite{hakeem2023security}. This network structure creates a challenging problem for implementing handover authentication, complicated by factors of multi-domain scenarios, frequent handovers, propagation issues, and scalable solutions. To address these challenges, research has proposed multiple 
solutions to incorporate 6G handover authentication for security and mobility. 

For instance, in \cite{kazmi2023security}, a variety of different architecture models are proposed for the inclusion of handover authentication in the 6G network. The first model includes lightweight clustering, with game-theory based decisions using AI and ML techniques. Frameworks of this type would reduce delays and signaling overhead but lack scalability and optimization for NTNs. The second model type uses proxy-based ring signatures. This technique enables secure handovers between mobile vehicles and edge nodes, however it primarily focuses on specific scenarios. A third model is proposed using blockchain-based methods and lightweight cryptographic techniques to offer secure handovers and reduce latency. However, the structure of this model overlooks complexity and compatibility issues. The last proposed model is built with a multi-connectivity architecture. This solution improves procedural efficiency and signal coverage but because it involves multiple base stations, increases the times required for base station discovery and random access.


\balance
    
\section{Conclusions}
\label{conclusion}
While 5G was a significant step forward in connectivity, it is vulnerable to a multitude of security vulnerabilities.  Many of the vulnerabilities manifest themselves during the initial stages of establishing a connection, including cell and network acquisition, registration, and authentication and key agreement.  Moreover, the densification of 5G networks and emergence of NTN communication modes has dramatically increased the rate of handovers, which are another large threat surface.  The emergence of inexpensive software defined radios has made it easy for an adversary to create a false base station and the lack of authentication makes it easy to trick a UE to attach or handover to such false base stations.

It is anticipated that 6G will address many of these concerns.   A simple step forward is to embed an authentication channel to the synchronization channels used for cell acquisition and handover.  Further improvements in the authentication and key agreement protocol as well as in the handover procedures are expected to further secure the network.   While it is too late for the changes advocated in this paper to be adopted into the 5G standard, it is recommended that 3GPP take into account these considerations as it standardizes 6G.

\section*{Acknowledgments}
This research was funded in part by the Department of Defense and by the National Science Foundation by way of the Center for Identification Technology Research (CITeR).




\bibliographystyle{IEEEtran}

\bibliography{library}

\begin{thebibliography}{10}
\providecommand{\url}[1]{#1}
\csname url@samestyle\endcsname
\providecommand{\newblock}{\relax}
\providecommand{\bibinfo}[2]{#2}
\providecommand{\BIBentrySTDinterwordspacing}{\spaceskip=0pt\relax}
\providecommand{\BIBentryALTinterwordstretchfactor}{4}
\providecommand{\BIBentryALTinterwordspacing}{\spaceskip=\fontdimen2\font plus
\BIBentryALTinterwordstretchfactor\fontdimen3\font minus \fontdimen4\font\relax}
\providecommand{\BIBforeignlanguage}[2]{{%
\expandafter\ifx\csname l@#1\endcsname\relax
\typeout{** WARNING: IEEEtran.bst: No hyphenation pattern has been}%
\typeout{** loaded for the language `#1'. Using the pattern for}%
\typeout{** the default language instead.}%
\else
\language=\csname l@#1\endcsname
\fi
#2}}
\providecommand{\BIBdecl}{\relax}
\BIBdecl

\bibitem{andrews2024}
J.~G. Andrews, T.~E. Humphreys, and T.~Ji, ``{6G} takes shape,'' \emph{IEEE BITS the Information Theory Magazine}, 2024.

\bibitem{guidotti2024}
A.~Guidotti, A.~Vanelli-Coralli, M.~E. Jaafari, N.~Chuberre, J.~Puttonen, V.~Schena, G.~Rinelli, and S.~Cioni, ``Role and evolution of non-terrestrial networks toward {6G} systems,'' \emph{IEEE Access}, vol.~12, pp. 55\,945--55\,963, 2024.

\bibitem{porambage2021roadmap}
P.~Porambage, G.~G{\"u}r, D.~P. Moya~Osorio, M.~Liyanage, A.~Gurtov, and M.~Ylianttila, ``The roadmap to {6G} security and privacy,'' \emph{IEEE Open Journal of the Communications Society}, vol.~2, 2021.

\bibitem{kazmi2023security}
S.~H.~A. Kazmi, R.~Hassan, F.~Qamar, K.~Nisar, and A.~A.~A. Ibrahim, ``Security concepts in emerging {6G} communication: Threats, countermeasures, authentication techniques and research directions,'' \emph{Symmetry}, vol.~15, no.~6, p. 1147, 2023.

\bibitem{hakeem2023security}
S.~A.~A. Hakeem, H.~H. Hussein, and H.~Kim, ``Security requirements and challenges of {6G} technologies and applications,'' \emph{Sensors}, vol.~22, no.~5, p. 1969, 2022.

\bibitem{cox2021book}
C.~Cox, \emph{An Introduction to 5G: The New Radio, 5G Network and Beyond}, 1st~ed.\hskip 1em plus 0.5em minus 0.4em\relax John Wiley \& Sons Ltd, 2021.

\bibitem{dahlman2021book}
E.~Dahlman, S.~Parkvall, and J.~Skold, \emph{5G NR: The Next Generation Wireless Access Technology}, 2nd~ed.\hskip 1em plus 0.5em minus 0.4em\relax Academic Press, 2021.

\bibitem{3GPP.TS.23.501}
3GPP, ``Services and system aspects; system architecture for the {5G} system {(5GS)}; stage 2,'' in \emph{3GPP TS 23.501 version 17.14.0}, Sep. 2024.

\bibitem{3GPP.TS.33.501}
------, ``Services and system aspects; security architecture and procedures for {5G} system,'' in \emph{3GPP TS 33.501 version 17.14.0}, Jun. 2024.

\bibitem{3GPP.TS.38.211}
------, ``Radio access network; {NR}; physical channels and modulation,'' in \emph{3GPP TS 38.211 version 17.9.0}, Sep. 2024.

\bibitem{3GPP.TS.38.331}
------, ``Radio access network; {NR}; radio resource control {(RRC)} protocol specification,'' in \emph{3GPP TS 38.331 version 17.10.0}, Sep. 2024.

\bibitem{3GPP.TS.38.321}
------, ``Radio access network; {NR}; medium access control {(MAC)} protocol specification,'' in \emph{3GPP TS 38.321 version 17.10.0}, Sep. 2024.

\bibitem{ludant2021sigunder}
N.~Ludant and G.~Noubir, ``{SigUnder}: A stealthy {5G} low power attack and defenses,'' in \emph{Proc. ACM Conference on Security and Privacy in Wireless and Mobile Networks (WiSec '21)}, 2021, pp. 250--260.

\bibitem{basin2018formal}
D.~Basin, J.~Dreier, L.~Hirschi, S.~Radomirovic, R.~Sasse, and V.~Stettler, ``A formal analysis of {5G} authentication,'' in \emph{Proc. ACM SIGSAC Conference on Computer and Communications Security (CSS'18)}, 2018, pp. 1383--1396.

\bibitem{cheng2022tracking}
Y.-C. Cheng and C.-A. Shen, ``A new tracking-attack scenario based on the vulnerability and privacy violation of {5G AKA} protocol,'' \emph{IEEE Access}, vol.~10, pp. 77\,679--77\,687, 2022.

\bibitem{bitsikas2021handover}
E.~Bitsikas and C.~Pöpper, ``Don’t hand it over: Vulnerabilities in the handover procedure of cellular telecommunications,'' in \emph{Proc. 37th Annual Computer Security Applications Conference (ACSAC'21)}, 2021, pp. 900--915.

\end{thebibliography}

\end{document}